\documentclass[fleqn,10pt]{wlscirep}
\usepackage{amsmath}
\usepackage{amssymb}
\usepackage{gensymb}
\usepackage{xargs}
\usepackage{dsfont}
\usepackage{color}
\usepackage{amsfonts}
\usepackage{bm}
\usepackage{graphicx}

\newcommand{\be}{\begin{displaymath}}
\newcommand{\ee}{\end{displaymath}}
\newcommand{\bn}{\begin{equation}}
\newcommand{\en}{\end{equation}}

\title{Enhanced target normal sheath acceleration  using colliding laser pulses}

\author[1,*]{J. Ferri}
\author[2]{E. Siminos}
\author[1]{T. F\"ul\"op}

\affil[1]{Department of Physics, Chalmers University of Technology, 41296 Gothenburg, Sweden}
\affil[2]{Department of Physics, Gothenburg University, 41296 Gothenburg, Sweden}

\affil[*]{julien.ferri@chalmers.se}

\begin{abstract}
  Laser-solid interaction can lead to the acceleration of protons to tens of MeV. Here, we show that a strong enhancement of this acceleration
  can be achieved by splitting the laser pulse to two parts of equal energy and opposite incidence angles.  
  Through the use of two- and three-dimensional Particle-In-Cell simulations, we find that the
  multi-pulse interaction leads to a standing wave pattern at the
  front side of the target, with an enhanced electric field and a
  substantial modification of the hot electron generation
  process. This in turn leads to significant improvement of the proton
  spectra, with an almost doubling of the accelerated proton
  energy and five-fold enhancement of the number of protons. 
  The proposed scheme is robust with respect to incidence
  angles for the laser pulses, providing flexibility to the scheme,
  which should facilitate its experimental implementation.
\end{abstract}
\begin{document}

\flushbottom
\maketitle

\thispagestyle{empty}

\section*{Introduction}

Proton acceleration due to the interaction of an ultraintense laser pulse with a thin solid target has been widely studied in the past two decades \cite{Daido2012,Macchi2013}, bringing the field closer to applications, such as proton therapy \cite{Bulanov2002, Bulanov2002b}, probing of electric fields \cite{Borghesi2007,Romagnani2008}, isochoric heating \cite{Patel2003} or  inertial confinement fusion \cite{Roth2001}. Although many different acceleration mechanisms were suggested, target normal sheath acceleration (TNSA) \cite{Wilks2001,Clark2000,Snavely2000, Maksimchuk2000} has been the most investigated method, thanks to a relatively simple experimental implementation and the moderate laser-intensity it requires. However, even with the continuous increase in the available laser power, the limited scaling of the maximum energy of the accelerated protons with the laser energy (which in general is, $E_{\text{max}}\propto I_0^{\alpha}$, with $\alpha<1$ depending on the laser pulse duration \cite{Fuchs2006, Robson2007, Zeil2010})  constitutes a major drawback for many applications.

Recent experiments proposed a way to improve the performance of the TNSA scheme by splitting the main laser pulse in two less-energetic pulses, incident on the target within a short time delay \cite{Markey2010,Scott2012,Brenner2014,Ferri2018}. It was suggested that such a multiple-pulse scheme could be used to produce mono-energetic features in the proton spectrum \cite{Robinson2007}. Later it was also shown experimentally that it could lead to an enhancement of the maximum proton energy \cite{Markey2010, Scott2012}, and that high laser-to-proton energy conversion efficiency could be reached \cite{Brenner2014}. However, the physical mechanism explaining these results relies on a judicious balance between plasma expansion on the front side and on the rear side. As a consequence, these experiments require an accurate control of both the time delay and the energy splitting between the two pulses, leading to a narrow range of parameters and limited energy enhancement. Additionally, in a recent experiment using an equal splitting of the laser energy into two femtosecond laser pulses, following each other within a controllable time delay and the same angle of incidence, no enhancement of the proton energy was obtained, even with a precise scanning of the time delay \cite{Ferri2018}.

In this article, we describe a modified TNSA scheme which consists in the splitting of a laser pulse in two pulses of equal energy which are incident on the thin solid target simultaneously, but with different angles of incidence. Based on two-dimensional (2D) simulations with the \textsc{epoch} Particle-In-Cell (PIC) code \cite{Arber2015}, and corroborated with three-dimensional (3D) simulations, we show that the interaction in the resulting standing wave leads to an increase in the peak value of the electric fields and substantial enhancement of the hot electron generation process at a constant laser energy. This in turn leads to a strong increase in the proton energy (from 8.5 to 14 MeV with a $45\degree$ angle and a high-contrast, 1.1~J laser) and proton number (by a factor of at least 5) with realistic laser parameters. Furthermore, we show that these conclusions remain valid for a large range of incidence angles for the laser pulses, including asymmetric configurations. Finally, we study the effect of the pre-plasma scale-length in the proposed scheme.

\section*{Results}
\subsection*{Enhancement of the proton energy}

\begin{figure}[t!]
\centering
\includegraphics[width=0.8\columnwidth]{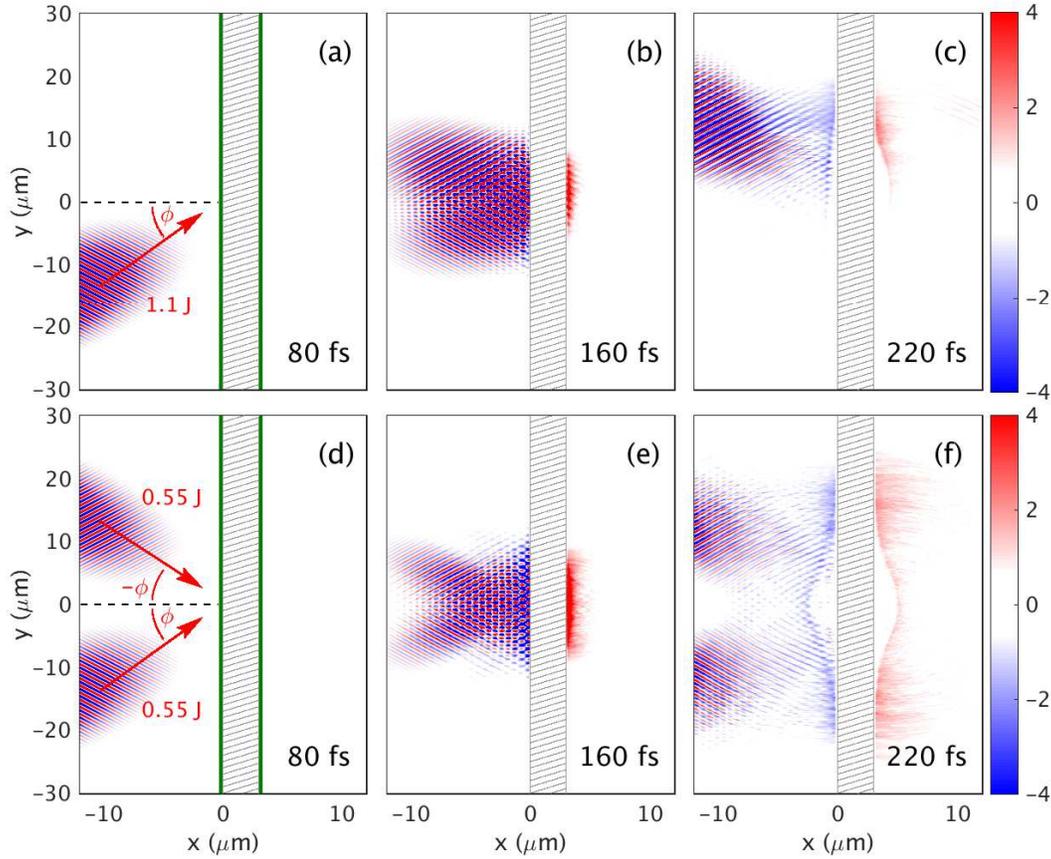}
\caption{Geometry of the two-pulse scheme. Map of the longitudinal electric field $E_x$ (blue-red, in TV~m$^{-1}$) in the ($x,y$) plane at different times in the reference case (a-c) and the two-pulse case (d-f). The position of the Al target is indicated by the dashed gray lines, and the thin proton layers are shown in green in (a) and (d). The incidence angle is $\phi=45\degree$.}
\label{scheme}
\end{figure}
To illustrate the scheme, an initial 1.1~J laser pulse is splitted in two sub-pulses of 0.55~J with Gaussian temporal and spatial profiles and  duration $\tau_0 = 38$~fs. These two $p$-polarized laser pulses are incident on a solid, $3~\mu$m-thick Aluminium (Al) target simultaneously and on the same perfectly overlapping $5~\mu$m spot (corresponding to an intensity $I_0 = 7\times10^{19}$W~cm$^{-2}$), but with opposite incidence angle $\phi$ and $-\phi$ with respect to the target normal in the $x,y$ plane. In the following, the case with one pulse containing the total 1.1~J energy will be referred to as the reference case, in contrast with the two-pulse case. We performed 2D-simulations of these cases with the \textsc{epoch} PIC code, see the Methods section for details. The proton species are assumed to originate from hydrogen-containing contaminants at the surface of the target, and are simulated with 20~nm-thick layers on the front and rear sides of the target. The scheme is illustrated with snapshots of the $E_x$ longitudinal electric field at different simulation times in Fig.~\ref{scheme} and compared with the reference case (Figs. \ref{scheme}(a)-(c)). Note that the two-pulse case (Figs. \ref{scheme}(d)-(f)) leads to a more symmetric mechanism for the generation of the rear-field. Furthermore, there is an enhancement of the rear field already at the beginning of the interaction (compare Figs.~\ref{scheme}(b) and \ref{scheme}(e)).

\begin{figure}[t!]
\centering
\includegraphics[width=0.8\textwidth]{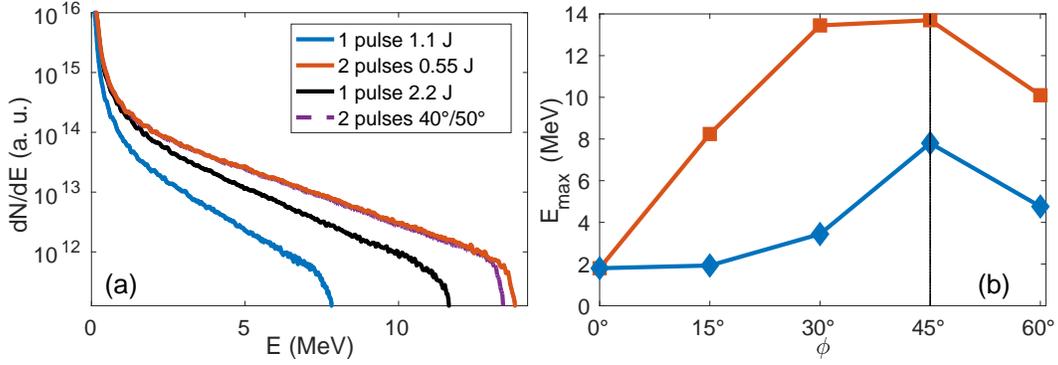}
\caption{Enhancement of the proton production in the two-pulse case. (a) Proton spectra (dN/dE) after 700~fs in the reference case (blue), the two-pulse case (red) and cases with one 2.2~J laser pulse (black) and with two 0.55~J laser pulses but slightly asymmetric incidence angle ($\phi_1 = 40\degree$ and $\phi_2 = -50\degree$, dashed purple). (b) Dependence of the maximum proton energy $E_{\text{max}}$ (in MeV) as a function of the incidence angle $\phi$. For the two-pulse case, the angles are $\phi$ and $-\phi$. The black dashed lines indicate $\phi=45\degree$ on which we mainly focus our analysis.}
\label{prot}
\end{figure}

The benefit of splitting the laser pulse in two is shown in Fig.~\ref{prot}(a), which presents the proton spectra at the end of the simulation ($t=700$~fs), in the reference case, the two-pulse case and a case with a pulse with double the energy for an incidence angle $\phi=45\degree$. Clearly, the two-pulse scheme leads to a strong increase in the proton energy, with maximum energy $E_{\text{max}}$ increasing from 7.8~MeV to 13.7~MeV when compared with the reference case, i.e.~an increase of approximately 80\%. Even compared with a case containing twice the total energy, i.e.~2.2~J, in a single laser pulse, there is a $\sim20$\% improvement in the two-pulse case, see the black curve in Fig.~\ref{prot}(a). Moreover, when integrating the spectra, the number of protons above 1~MeV is multiplied by a factor of $>5$ in the two-pulse case.

The flexibility of our scheme with respect to the incidence angle is explored in Fig.~\ref{prot}(b), in which the maximum proton energies for the reference and the two-pulse cases are shown for a wide range of incidence angles. In the reference case, a clear peak of the proton energy is observed at $45\degree$, corresponding to  the maximum efficiency of the 
vacuum heating for these angles \cite{Gibbon1992}. In the single-pulse case, the TNSA mechanism is quickly degraded with decreasing angles and is inefficient for small incidence angles. In contrast, in the two-pulse case, the mechanism is sustained for a wider range of angles, with small variation of $E_{\text{max}}$ for $\phi$ in the range $30\degree$--$60\degree$. As a consequence, the proton energy enhancement is more pronounced for angles of incidence different than the one-pulse optimal $45\degree$, reaching $300\%$ for $\phi=30\degree$.
 
In this paper we mainly explore a symmetric double-pulse scenario, but the scheme can be extended to asymmetric incidence angles. The practical advantage with the asymmetry is that the reflected laser pulses will not aim straight back for the optics, and therefore the experimental implementation of the scheme will be easier. In Fig.~\ref{prot}(a), we show a proton spectrum corresponding to a two-pulse asymmetric case, with $\phi_1 = 40\degree$ and $\phi_2 = -50\degree$. Introducing this asymmetry keeps the results largely unchanged, with the maximum proton energy degraded by a mere 3\%. This indicates that the main effect can be detected even in an experiment with e.g.~a tilted target in the direction orthogonal to the propagation plane of the laser pulses.

\subsection*{Modification of the vacuum heating mechanism}

To understand the difference in the acceleration mechanism between the single and multiple-pulse cases we need to study the differences in the electromagnetic field
configurations close to the target surface and the ways in which these affect absorption. 
In our simulations, we observe two different types of electron populations: most of the electrons stay in the first wavelength of the standing wave in front of the target, meaning that they are accelerated through the vacuum heating mechanism~\cite{Brunel1987, Brunel1988} over a single wavelength of the electromagnetic wave. This is coherent with our conditions, i.e. oblique incidence and flat target profile. By contrast, a small proportion of the population is ejected further away and is usually re-injected in the target with slightly higher average momentum. For this population, additional stochastic heating is present, but this effect remains marginal in both cases (as also noted in Ref.~\cite{Yogo2016}).

\begin{figure}[t!]
\centering
\includegraphics[width=0.8\columnwidth]{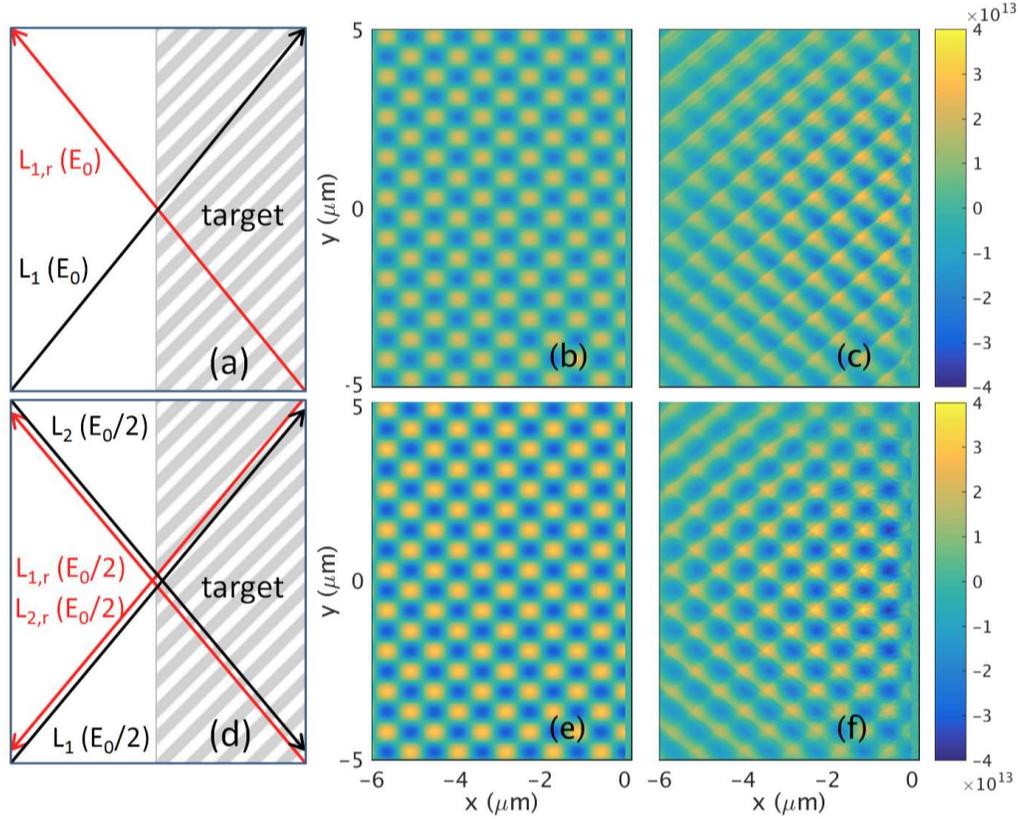}
\caption{Fields in the multi-pulse interaction. Panels (a)-(c) correspond to the reference simulation and panels (d)-(f) to the two-pulse case. (a),(d) Scheme of the multi-pulse interaction showing the incoming laser pulses $L_1$ and $L_2$ and the reflected laser pulses $L_{1,\text{r}}$ and $L_{2,\text{r}}$. (b),(e) Map of the analytical longitudinal electric field $E_x$ (blue-yellow, in V~m$^{-1}$) in the $(x,y)$ plane obtained from equations (\ref{eq1}) and (c),(f) snapshot of the longitudinal electric field $E_x$ (blue-yellow, in V~m$^{-1}$) in front of the target after 155~fs at the peak interaction.}
\label{multipulse}
\end{figure}

For simplicity, and since we are mostly interested in qualitatively understanding how the two cases differ in terms of the vacuum heating mechanism, we consider the target as a perfect conductor, i.e.~the laser pulses are perfectly reflected on the surface of the target. One could then describe the incoming and reflected fields as an effective multi-pulse configuration in the half-space in front of the target. Letting $N$ be the number of effective (i.e. incident plus reflected) pulses, we have $N=2$ pulses of intensity $I_0$ in the reference case (see Figs.~\ref{multipulse}(a)-(c)) and $N=4$ pulses of intensity $I_0/2$ in the two-pulse case (see ~Figs.~\ref{multipulse}(d)-(f)). The key benefit of a multi-pulse system is that the field amplitude scales as $\sqrt{N}$, as observed in Ref.~\cite{Bulanov2010}, where three-dimensional solutions for the field of focused pulses were obtained. For the sake of simplicity, we here consider the case of $N$ plane waves, with vector-potentials given by $A_n = A_0\sqrt{2/N}\cos(\omega_0t-k_0\cos(\phi_n)x-k_0\sin(\phi_n)y)$, where $\phi_n$ is the angle of incidence for the $n^{th}$ pulse, $\omega_0 = 2\pi c/\lambda_0$ and $k_0 = \omega_0/c$, with $c$ the speed of light. For $p$-polarized waves, the electric field $E_x$ and magnetic field $B_z$ in front of the target can then be written:
\begin{align}
 E_{x,2} &= -2\sin\phi E_0\sin(t-y\sin\phi)\cos(x\cos\phi), & B_{z,2} &= 2B_0\sin(t-y\sin\phi)\cos(x\cos\phi), \nonumber \\
 E_{x,4} &= 2\sqrt{2}E_0\sin\phi\cos t\sin(y\sin\phi)\cos(x\cos\phi), & B_{z,4} &= -2\sqrt{2}B_0\sin t \cos(y\sin\phi)\cos(x\cos\phi),
 \label{eq1}
\end{align}
where time has been normalized to $\omega_0^{-1}$ and coordinates to $k_0^{-1}$, $E_0 = \omega_0A_0$, $B_0=E_0/c$, the incident angles are assumed to be symmetric with respect to the $x$ axis and $E_{x,N}$ denotes
the field in the case of $N$ effective pulses. 
The peak field is increased by $\sqrt{2}$ in the two-pulse case ($N=4$) compared to the reference case ($N=2$). 
It can also be seen from Eq.~\ref{eq1} that in the two-pulse case a standing wave is obtained in front of the target, where the fields are $2\pi/\omega_0$ periodic in time. These fields are shown in Figs.~\ref{multipulse}(b) and \ref{multipulse}(e), and can be compared with the fields obtained from the simulations, at the time of interaction of the peak of the laser pulse with the target (and at the peak value of the fields for the two-pulse case) (Figs.~\ref{multipulse}(c) and \ref{multipulse}(f)). Note, that the analytical and simulation fields are similar close to the focal spot, particularly in the two-pulse case. In the reference case, the field symmetry is slightly changed by the modification of the reflected pulse (partial absorption and high-harmonic generation).

\begin{figure}[t!]
\centering
\includegraphics[width=\textwidth]{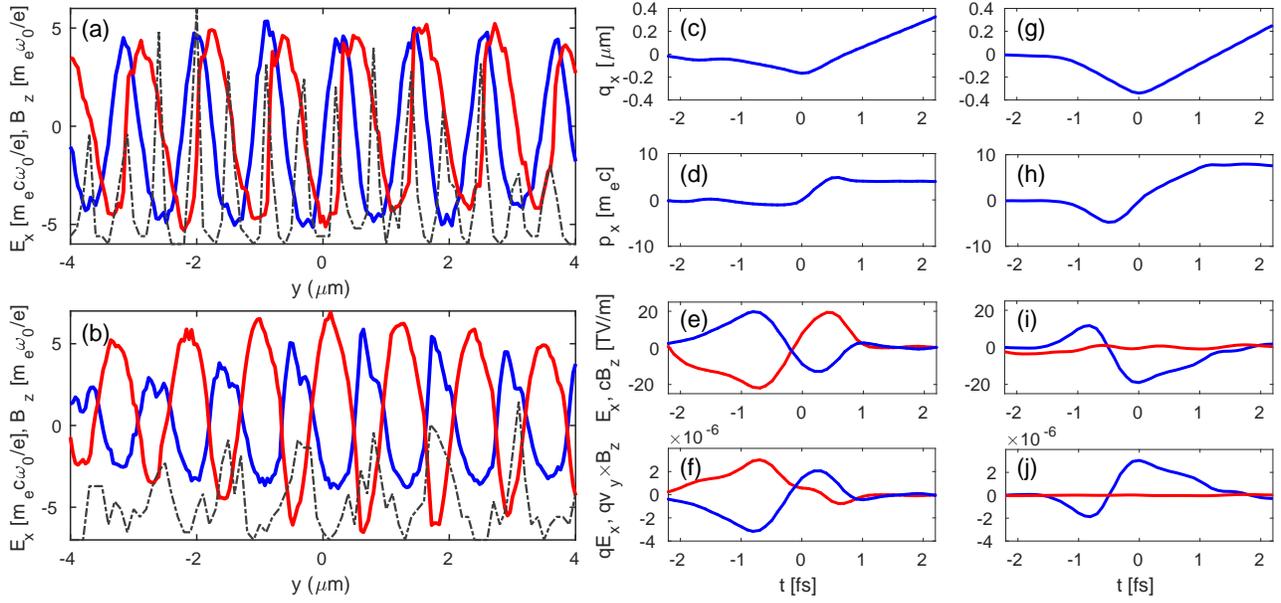}
\caption{Modification of the fields generating the hot electrons. (a) (resp~(b)) $E_x$ (blue) and $B_z$ (red) fields on the target surface at $t=150~$fs in the two-pulse case (resp.~reference case). These fields are indicated in normalized units, i.e. $m_\text{e} c \omega_0/e$ for $E_x$ and $m_\text{e} \omega_0/e$ for $B_z$, with $m_e$ and $e$ respectively the electron mass and elementary charge. The dashed black lines indicate the distribution of the initial $y$ position on the target surface for the electrons reaching 400~keV during the simulation. For the reference case, this distribution is plotted in the frame moving with a $c\sin\phi$ velocity along $y$ to take into account the drifting of the fields. Panels (c)-(f) correspond to the reference simulation and panels (g)-(j) to the two-pulse case. (c),(g) Longitudinal position $q_x$, (d),(h) longitudinal momentum $p_x$, (e),(i) $E_x$ (blue) and $B_z$ (red) fields and (f),(j) $qE_x$ (blue) and $qv_y\times B_z$ (red) forces experienced as a function of time for an accelerated electron in the two different cases. $t=0$ corresponds to the time at which the particles reach their minimum longitudinal position.
}
\label{ExBz}
\end{figure}

Although the vacuum heating mechanism has been the subject of
intense study at increasing levels of sophistication~\cite{Gibbon1992,Geindre2006,Debayle2013,Mulser2012}, 
here we are merely interested in developing a qualitative understanding 
of the differences between the single- and double-pulse cases. We will thus 
only consider the simple ``capacitor'' model 
originally formulated by Brunel~\cite{Brunel1987}.
For a perfect conductor, it yields for the average power absorbed per laser cycle,
  $P_\text{a} \propto E_\text{d} \left[\left(1+e^2\,E_{\text{d}}^2/m_\text{e}^2c^2\omega_0^2\right)^{1/2}-1\right]$,
where the capacitor field $E_{\text{d}}$ is related to the driving 
longitudinal field by $E_{\text{d}}=2E_0\sin\phi$ for the single pulse case
and by $E_{\text{d}}=2\sqrt{2}E_0\sin\phi$ for the two-pulse case.
In general, the absorbed power would be expected
to scale as $P_\text{a}\propto N\,E_0^2$ in a configuration with $N$ effective pulses. 
However, for focused pulses the field enhancement and 
the associated increase in absorption is \emph{localized} close to the focus.

The field enhancement in the focus of a $N$-pulse system is however not sufficient to completely explain the difference between the two schemes. The capacitor model has been derived by assuming a longitudinal standing wave field in front of the target, while it neglects the effect of the $\mathbf{v}\times \mathbf{B}$ force on the electrons. As already noted in Ref.~\cite{Brunel1987} these approximations are only valid  in a two-pulse but not in a single-pulse case. This becomes clear if we take a closer look at the change of relative phase between the $E_x$ and $B_z$ fields. As noted in Ref.~\cite{Brunel1988}, the positions of the maxima of the $|E_x|$ field on the target surface coincide with nodes for the $B_z$ field in the two-pulse case, as shown in Fig.~\ref{ExBz}(a). Therefore, in the two-pulse case, the electron motion is mainly determined by the quiver motion in the $E_x$ field as the $\mathbf{v}\times \mathbf{B}$ force is negligible, and one can see that the electrons reaching high energy originate from very localized regions on the target surface corresponding to the maxima of $|E_x|$ (black dashed line). However, in the single-pulse case $B_z$ is not zero at these positions (see Fig.~\ref{ExBz}(b)), so the $\mathbf{v}\times \mathbf{B}$ force acts on the electrons, and largely inhibits the force due to $E_x$ \cite{Geindre2006}. 
This interpretation is also corroborated by particle-tracking simulations presented in Figure \ref{ExBz}(c-j), in which we compare the forces -- extracted from the PIC simulations -- applied on two electrons in the two cases. In both cases the particle is first pulled out of the target by the $E_x$ field, before being reinjected when the $E_x$ field reverses, eventually being able to escape into the target. In the one-pulse case, $E_x$ and $B_z$ have opposing phases (Fig.~\ref{ExBz}(e)), so that the $\mathbf{v}\times\mathbf{B}$ force mainly counteracts the force due to $E_x$ (Fig.~\ref{ExBz}(f)). However in the two-pulse case, the value of $B_z$ remains very low during
the whole acceleration of the electron (Fig. \ref{ExBz}(i)), and the  $\mathbf{v}\times\mathbf{B}$ force applied on the electron remains negligible compared to the force due to $E_x$ (Fig. \ref{ExBz}(h)). As a consequence,
when interacting with a $E_x$ field of similar value, the electron is pulled further away from the
target (compare Figs. \ref{ExBz}(c) and \ref{ExBz}(g)) and reaches a higher energy in the two-pulse case than in the
one-pulse case (compare Figs. \ref{ExBz}(d) and \ref{ExBz}(h)), as it is not influenced by the inhibiting $\mathbf{v}\times\mathbf{B}$ force.

The dependence of the proton energy on the incidence angle, shown in Fig.~\ref{prot}(b), can also be understood considering the relative effect of $E_x \propto E_0 \sin\phi$ and $v_y \times B_z\propto E_0 B_0\cos\phi$, where $B_0\sim E_0/c$ is the single pulse peak magnetic field. For the single-pulse case, $E_x$ then decreases when the angle of incidence gets smaller while $v_y \times B$ increases, leading to quick deterioration of hot electron generation, and thus to a clearly optimum angle of incidence at $\phi=45\degree$. This is not the case in the two-pulse setup, for which $B_z$ is zero at the maxima of $|E_x|$ independently of the angle of incidence. 

\begin{figure}[t!]
\centering
\includegraphics[width=\textwidth]{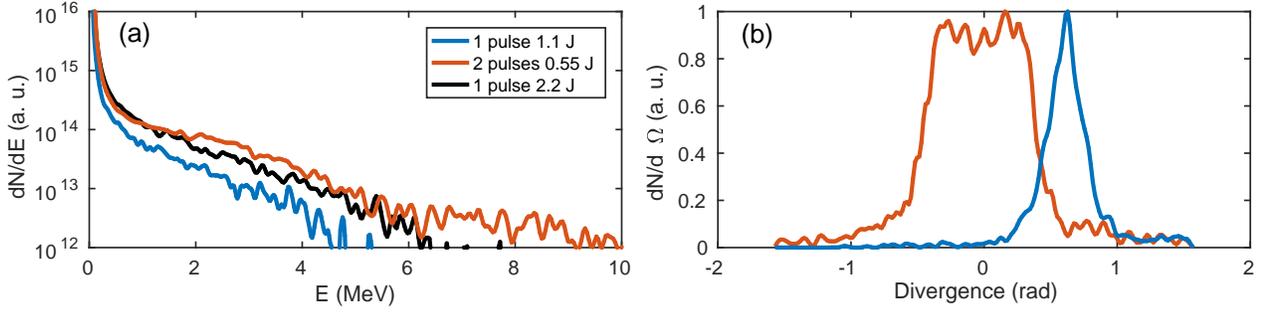}
\caption{Enhanced hot electron generation in the two-pulse case. (a) Energy spectra $dN/dE$ and (b) divergence $dN/d\Omega$ of the hot electron beam after 160~fs (interaction of the peak of the laser pulse with the target) for the reference case (blue), the two-pulse case (red) and a case with only one pulse but twice the total energy (black). $\Omega$ defines the angle with the $x$ direction in the (x,y) plane.}
\label{elec}
\end{figure}

The above effects demonstrate that, compared to the single-pulse case, the two-pulse configuration leads to an increased number and energy of the vacuum-accelerated electrons. This is confirmed in the electron spectra presented in Fig.~\ref{elec}(a), taken from the simulation during the interaction of the peak of the laser pulse with the Al target, where we can see that a larger number of more energetic hot electrons are generated in the two-pulse case: Maxwellian fits yield temperatures of 1.05~MeV and 1.40~MeV, respectively, in the one-pulse and two-pulse cases. We also plot the electron spectra obtained with a single laser pulse with twice the energy (2.2~J, black). In this configuration, the peak field is the same as in the two-pulse case with 1.1~J of total energy, but the temperature of the spectra is still inferior, showing the importance of the two-pulse geometry where the $\mathbf{v}\times \mathbf{B}$ force does not cancel $E_x$.
The higher hot electron energy density will in turn impact the electrostatic fields generated on the rear of the target (shown in Fig.~\ref{scheme}(e)), for the benefit of the TNSA process. The increase in the proton energies agrees with the standard estimate for the rear sheath field $E_{x,\text{sheath}} \propto \sqrt{n_\text{H}T_\text{H}}$~\cite{Mora2003}, with $n_\text{H}$ and $T_\text{H}$ the hot electron density and temperature obtained from Fig.\ref{elec}(a) (in the simulations $n_\text{H}$ is increased by a factor $\sim2$ between the single-pulse and two-pulse cases).

Note that while we are interested in the modification of the hot electron population and of the proton acceleration in this paper, the modification of the standing-wave fields in the two-pulse scheme might also affect the backward-accelerated electrons that have been investigated in previous studies\cite{Mordovanakis2009,Morrison2015,Orban2015}.

We also note a change in the divergence of the electron
beam, shown in Fig. \ref{elec}(b). In the single-pulse scheme, the
electrons are preferentially accelerated in the direction of
the incoming laser pulse, with a peak of the distribution
at $\sim 45 \degree$ . As expected in the symmetric two-pulse mechanism, the hot electron divergence is centered along the $x$-axis.
However in this case, the divergence of the electron beam
is increased, so the rear TNSA fields should decrease faster
with increasing thickness of the target in the two-pulse
scheme.

Finally, the efficiency of the two-pulse scheme does not depend on the relative phase between the two laser pulses, provided that the time difference is much shorter than the pulse duration. This is due to the fact that the standing wave pattern is determined solely by the reflection conditions. This was verified in simulations (not shown). Similarly, the mismatch between the focal position of the laser pulses should not exceed the spot size, in order not to degrade the performance of the scheme.

\subsection*{Effect of preplasma}

As the efficiency of the scheme relies on the vacuum acceleration mechanism, taking into account the effect of a preplasma will affect the performance of the scheme. We therefore run a few additional simulations with preplasma of different lengths, with exponential profiles ranging from $0.01~n_\text{c}$ to $100~n_\text{c}$ and a scale length $L_\text{G}$ (see section Methods for further details). The results are presented in figure \ref{grad}(a), and show an increase of the proton energy for both the one-pulse and two-pulse cases when introducing a preplasma. Both cases exhibits an optimal case for the maximum proton energy when $L_\text{G}\sim0.6~\mu$m. While the gap between the one-pulse and two-pulse cases closes when $L_\text{G}$ increases, the two-pulse case remains more efficient for all the preplasma tested, and in particular, $E_{\text{max}}$ is still increased by more than 25\% in the optimal case of $L_\text{G}=0.6~\mu$m. This indicates that our method, which is very efficient when applied to ultra-high contrast laser systems using for example double plasma mirror~\cite{Kapteyn1991,Doumy2004}, will still be of interest for lower contrast facilities, although with lower gain compared to a single-pulse scheme. The two-pulse scheme might be of particular interest when using ultrathin foils, for which ultrahigh contrast is required~\cite{Neely2006,Ceccotti2007}.
Note that we only consider femtosecond laser durations in this study, since at present there appears to be no picosecond laser pulses with sufficiently high contrast to prevent early expansion of the target.

\subsection*{Energy scaling}

\begin{figure}[t!]
\centering
\includegraphics[width=\textwidth]{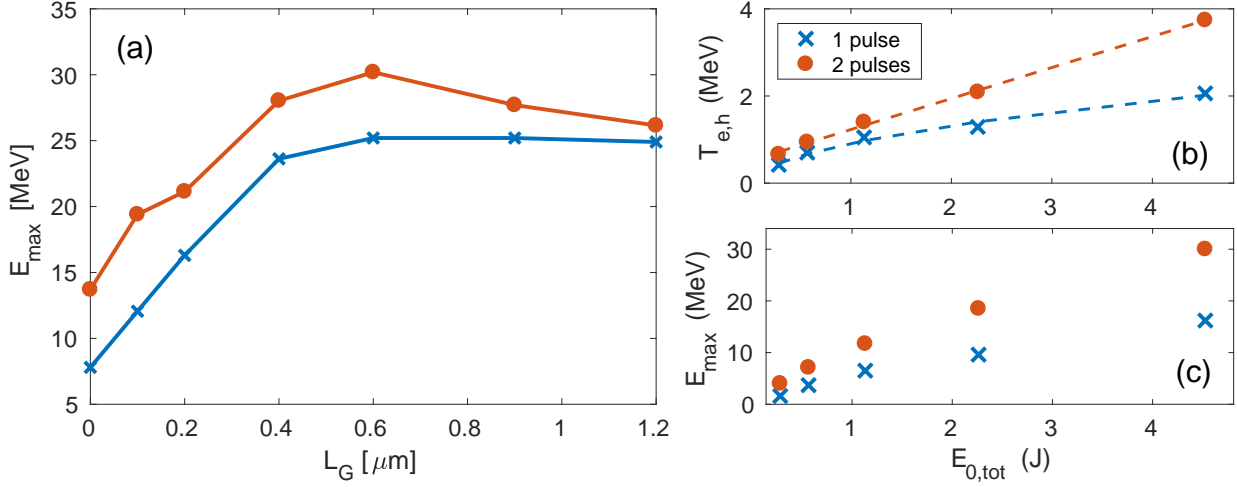}
\caption{Influence of the presence of a preplasma and energy scaling. (a) Maximum proton energy $E_{\text{max}}$ (MeV) as a function of the preplasma scale length $L_\text{G}$ (in $\mu m$) in the one-pulse (blue crosses) and two-pulses cases (red circles). $L_\text{G}=0$ corresponds to the reference cases with a sharp density profile. (b) Hot electron temperature $T_{\text{e,h}}$ (MeV) and (c) maximum proton energy $E_{\text{max}}$ (MeV) as a function of the total laser energy $E_{0,\text{tot}}$ (J).}
\label{grad}
\end{figure}

Figure \ref{grad}(b) and (c) presents results of simulations with a flat target profile and a $45\degree$ angle of incidence, but with a varying laser energy.
The hot electron temperature $T_{\text{e,h}}$ in the two-pulse
  case exhibits an almost linear dependence with the total laser
  energy $E_{0,\text{tot}}$, see Fig.~\ref{grad}(b) (fit in dashed red
  line). This is much more favorable than classical vacuum heating
  with one pulse, which scales as $E_{0,\text{tot}}^{1/2}$ (fit in dashed blue
  line), in agreement with Ref.~\cite{Gibbon1992}. The linear scaling
  is advantageous for higher energy laser systems, as the scaling for
  $T_{\text{e,h}}$ in the case of resonant absorption in the presence of a
  preplasma is expected to be weaker than
  $E_{0,\text{tot}}^{1/2}$~\cite{Forslund1977, Estabrook1978}. As the dependence of the TNSA rear fields on $T_{\text{e,h}}$ is relatively weak, the proton energy enhancement ratio remains approximately constant accross a range of intensities, (cf Fig.~\ref{grad}(c)).

For pulses with energy $E_{0,\text{tot}}>10$\,J we have found that the proton layer on the rear side starts to entirely disconnect from the bulk of the target during the acceleration.
As a consequense, the escaping accelerated proton layer stops interacting effectively with the accelerating field and the maximum proton energies start to saturate (both in the single-pulse and the two-pulse cases) and become lower than expected by the linear scaling. The onset of this effect depends on the layer parameters and target composition. While adjusting target parameters could allow exploiting the two-pulse scheme for even higher laser energies, such a parametric study is beyond the scope of this work.

\begin{figure}[t!]
\centering
\includegraphics[width=\textwidth]{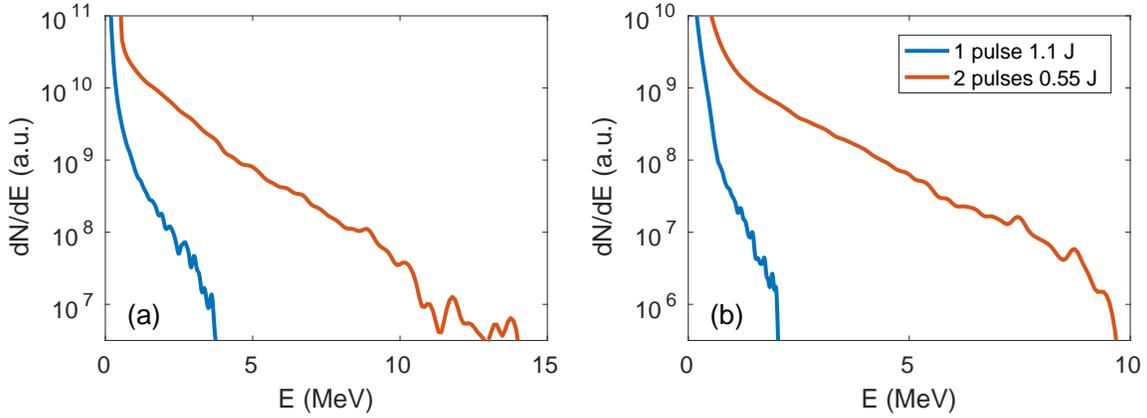}
\caption{Results in 3D simulations. (a) Electron and (b) proton spectra $dN/dE$ obtained in the 3D simulations after respectively 140~fs (interaction of the peak of the laser pulse with the target) and 400~fs (end of the simulation) in the single-pulse (blue) and two-pulse (red) cases.}
\label{3D}
\end{figure}

\subsection*{3D simulations}
In this section, we present proof-of-principle simulations showing that the proposed setup remains valid in a three-dimensional geometry. To this end, we performed simulations using the 3D version of the \textsc{EPOCH} PIC code, with parameters detailed in the Methods section. Physical parameters were mainly left unchanged, except that we reduced the thickness of the target to $1.5~\mu$m and reduced the aluminium target density so that the simulations would be manageable. For the same reason, we used an incidence angle of $30\degree$, which allows for a reduced box size.

Results concerning the hot electron and proton distributions in these 3D simulations are shown in Figure~\ref{3D}. As seen previously in the 2D simulations, splitting the pulse leads to an increase of the hot electron temperature (from 1.2~MeV to 1.5~MeV, cf Fig.~\ref{3D}(a)). This in turn leads to a much higher proton energy in the two-pulse case than in the single-pulse case ($E_{\text{max}}=9.6$~MeV compared with 2~MeV, cf Fig.~\ref{3D}(b)).
In the 3D simulations, $E_{\text{max}}$ is increased by a factor 4.7 between the single-pulse and two-pulse cases, while this factor was 3.9 in the 2D simulations (cf Fig. \ref{prot}(b)).
These values should not be compared quantitatively to  the ones obtained in the 2D simulations, due to the modifications of several physical and numerical parameters and to the fact that energies are generally lower in 3D simulations (as the rear field decreases faster when adding a dimension). However, the qualitative agreement between the 2D and 3D results confirms that our scheme is valid in a three dimensional geometry.

\section*{Discussion}

In conclusion, we have proposed a modification of the TNSA scheme aiming to enhance the proton energy at a constant laser energy. By splitting a main laser pulse into two laser pulses, simultaneously incident on a target at different angles, we can access a different energy repartition of the electromagnetic fields on the front side of the target, which in turn can generate a higher number of hotter electrons. This leads to a significant enhancement of the proton energy ($\sim 80$\% for a constant laser energy), and the number of protons (increase by a factor $>5$) in the case of a sharp density profile. This might be used in ultrahigh contrast laser facilities and is of very high interest for proton acceleration from ultrathin foils. In the presence of a preplasma generated by lower contrasts, the two-pulse scheme remains more efficient although with lower gain. In particular, we found an optimal case for the proton energy using both the two-pulse case and a preplasma, with 25\% of increase compared with a single-pulse scheme.
The robustness of this scheme, in particular with respect to the incident angles, makes it a strong candidate for future experiments, as setups allowing for spatial and temporal separation of a laser pulse have already been implemented \cite{Aurand2015, Aurand2016}. This scheme then potentially allows for higher performance of the TNSA mechanism than what can be obtained through temporal pulse shaping, and could possibly be further optimized by splitting the laser pulse into a larger number of sub-pulses.

\section*{Methods}

\subsection*{Particle-in-cell simulations}
Simulations were performed using the 2D version of the \textsc{epoch} PIC code. We used a $44\times 80~\mu$m$^2$ box with a $10$~nm resolution, and 50 particles per cell for the electron and ion species forming the $3~\mu$m-thick target. The proton species, assumed to originate from hydrogen-containing contaminants at the surfaces of the target, are simulated with 20~nm-thick layers at a $100n_\text{c}$ density on the front and rear sides of the target, using 1000 particles per cell. The target is initially fully ionized, with initial densities $n_\text{i} = 50n_\text{c}$ for the Al$^{13+}$ ions and $n_\text{e} = 13n_\text{i}$ for the electrons, where $n_\text{c}$ is the critical density, and the physical ion-to-electron mass ratio $m_\text{i}/m_\text{e} = 1836\times 27$ is used.

In the reference case, we employ a single 1.1~J Gaussian pulse, with a $38$~fs (FWHM) duration, $5~\mu$m focal spot, $0.8~\mu$m wavelength, leading to an intensity $7\times10^{19}$~W~cm$^{-2}$. The laser pulse is incident on target with an incidence angle $\phi$. In the two-pulse case, the energy of each pulse is divided by two in order to conserve the total laser energy, and the two pulses are incident on target with respective angle of incidence $\phi$ and $-\phi$, while all other parameters are kept unchanged.

When needed, the preplasma is modeled by a mixture of protons and electrons following an exponential density profile $n_\text{e} \propto \exp(x/L_\text{G})$, with $L_\text{G}$ the scale length of the preplasma. The preplasma expands from $0.01~n_\text{c}$ to $100n_\text{c}$ at the target surface in $x=0$, so that the length of the preplasma $l_\text{p}$ depends on $L_\text{G}$ as indicated in table \ref{tab1}. Besides, the focal point and position of crossing of the laser pulses in the two-pulse scheme is shifted to the position of the relativistic critical density $\sqrt{1+a_0^2}n_\text{c}$.

\begin{table}[ht]
\centering
   \begin{tabular}{| c || c | c | c | c | c | c |}
     \hline
       $L_\text{G}~[\mu$m]& 0.1 & 0.2 & 0.4 & 0.6 & 0.9 & 1.2\\ \hline
       $l_\text{p}~[\mu$m] & 0.9 & 1.8 & 3.7 & 5.5 & 8.3 & 11.1 \\ \hline
   \end{tabular}
   \caption{\label{tab1}Scale length $L_\text{G}$ and total length $l_\text{p}$ of the preplasma in the different simulations.}
\end{table}

3D simulations were also performed using the \textsc{epoch} PIC code. We used a $32\times 32\times 24~\mu$m$^3$ box size, divided in $1600\times 1280\times 960$ cells respectively in the $x$ (longitudinal), $y$ (laser polarization) and $z$ direction. We used 5 particles per cell for the electron and aluminium ion species, and 20 particles per cell for the proton species. The thickness and density of the aluminium target were respectively decreased to $1.5~\mu$m and $n_\text{i} = 10n_\text{c}$, while the proton layers were unchanged. Laser parameters were also kept unchanged, with polarization and incidence angles in the $x,y$ plane. These simulations were performed at a $30\degree$ angle of incidence, in order to limit the required box size.

\section*{Data availability}
The datasets supporting the findings of this study are available from the corresponding author on reasonable request.

\section*{Acknowledgments}

The authors would like to acknowledge fruitful discussions with
L~Gremillet, I~Thiele, J~Martins, L~Yi and the \textsc{PLIONA}
team. This work was supported by the Knut and Alice Wallenberg
Foundation, the European Research Council (ERC-2014-CoG grant 64712)
and by the Swedish Research Council, Grant No. 2016-05012. The
simulations were performed on resources at Chalmers Centre for
Computational Science and Engineering (C3SE) provided by the Swedish
National Infrastructure for Computing (SNIC, Grants SNIC 2017/1-484,
SNIC 2017/1-393, SNIC 2018/1-43, SNIC 2018/3-297).

\section*{Author contributions statement}
J.F. conceived the idea and performed the simulations in collaboration with E.S. J.F. and E.S. developed the theoretical interpretation. J.F., E.S. and T.F. discussed the findings and contributed to the writing of the manuscript.

\section*{Competing interests}
The authors declare no competing interests. 

\end{document}